\RequirePackage{fix-cm}
\documentclass[twocolumn]{svjour3}       
\smartqed
\usepackage{graphicx}
\usepackage{amsmath,amsfonts,amssymb}
\usepackage{bm}
\usepackage{dcolumn}

\begin{document}

\title{Microscopic origins of shear stress in dense fluid-grain mixtures}
\author{Donia Marzougui   \and  Bruno Chareyre   \and     Julien Chauchat}
\institute{D. Marzougui \at
              Univ. Grenoble Alpes, 3SR, F-38000 Grenoble, France and\\ CNRS, 3SR, F-38000 Grenoble, France\\
              \email{donia.marzougui@3sr-grenoble.fr}           
           \and
           B. Chareyre \at
              Univ. Grenoble Alpes, 3SR, F-38000 Grenoble, France and\\ CNRS, 3SR, F-38000 Grenoble, France\\
              \email{bruno.chareyre@3sr-grenoble.fr}           
           \and
           J.Chauchat \at
	      Univ. Grenoble Alpes, LEGI, F-38000 Grenoble, France and\\ CNRS, LEGI, F-38000 Grenoble, France\\
              \email{julien.chauchat@grenoble-inp.fr}           
}

\date{Received: \today / Accepted: date}

\begin{abstract}
A numerical model is used to simulate rheometer experiments at constant normal stress on dense suspensions of spheres.
The complete model includes sphere-sphere contacts using a soft contact approach,
short range hydrodynamic interactions defined by frame-invariant expressions of forces and torques in the lubrication approximation,
and drag forces resulting from the poromechanical coupling computed with the DEM-PFV technique.
Series of simulations in which some of the coupling terms are neglected highlight the role of the poromechanical coupling in the transient regimes.
They also reveal that the shear component of the lubrication forces, though frequently neglected in the literature, has a dominant effect in the volume changes.
On the other hand, the effects of lubrication torques are much less significant.

The bulk shear stress is decomposed into contact stress and hydrodynamic stress terms
whose dependency on a dimensionless shear rate - the so called viscous number $I_v$ - are examined. Both contributions are increasing 
functions of $I_v$, contacts contribution dominates at low viscous number ($I_v<$0.15) whereas lubrication contributions are
dominant for $I_v>$ 0.15, consistently with a phenomenological law infered by other authors.
Statistics of microstructural variables highlight a complex interplay between solid contacts and hydrodynamic interactions. In contrast with a popular idea, the results suggest that lubrication may not necessarily reduce the contribution of contact forces to the bulk shear stress.
The proposed model is general and applies directly to sheared immersed granular media in which pore pressure feedback plays a key role (triggering of avalanches, liquefaction).

\end{abstract}



\maketitle

\section{Introduction}
\label{intro}

Dense suspensions of particles immersed in a viscous fluid are ubiquitous in natural phenomenon, such as sediment transport or debris flows, and in numerous industrial applications such as form filling with fluid, concrete in civil engineering or slurry transport in petroleum industries. The understanding of dense suspension  rheology has lead to an important research effort over the past decades \cite{Bagnold1954,frankel1967viscosity,brady1988stokesian}. The complexity of this problem arises from its two-phase nature involving a fluid phase (continuous) and a particulate phase (discrete) for which particle-particle interactions and fluid-particle interactions contribute to the behavior of the system in the dense limit. 

Classical rheometer experiments impose simple shear of the suspensions at constant volume (type I). In such case, the interpretation of the shear stress in terms of effective viscosity $\eta_{e}$ suggests that $\eta_{e}$ diverges at high solid fraction ($\phi\approx0.6$ for spheres) \cite{stickel2005}. Rheometer experiments at constant normal stress $P$ have been performed only recently \cite{boyer2011unifying} (type II). Under such conditions the volume of the suspension is free to change as a response to the imposed shearing. On this basis, a description of dense suspensions has been proposed, which unifies classical suspension rheology, described in terms of a shear and normal effective viscosity which depend on the solid fraction, and the dense granular flow rheology, described in terms of shear to normal stress ratio ($\mu$) and solid fraction ($\phi$). For this purpose a dimensionless shear rate was introduced, the so called  \textit{viscous number} ($I_v$), controlling both frictional and viscous 
contributions to the shear stress. An advantage of such a visco-plastic vision is that the behaviour of the suspension in the very dense limit is not associated with a divergence of the viscosity. Instead, the shear to normal stress ratio reaches a constant value corresponding to a Coulomb-type bulk friction. 

From an analytical point of view, the rheology of suspensions has been studied since the beginning of the 20st century. Einstein in 1905 \cite{einstein1905molekularkinetischen} derived the effective viscosity of a dilute suspension based on long range hydrodynamic interactions. Frankel and Acrivos \cite{frankel1967viscosity} proposed another derivation in the limit of dense packings. The later was based on the so-called \emph{lubrication} approximation of the hydrodynamic interactions between nearly touching spheres. The lubrication terms are singular, they appear as pair interactions between particles and they diverge at the approach of contact.

The first discrete numerical simulations of particle suspension has been proposed in the framework of \emph{Stokesian Dynamics} (SD) \cite{bossis1984dynamic,brady1985rheology} using resistance and mobility matrices \cite{jeffrey1984forces}. This technique is able to quantitatively reproduce the divergence of effective viscosity for solid fraction approaching close packing. In the general framework of SD, the hydrodynamic forces include both long range and short range interactions which are defined independently, the later are the diverging terms as found in the lubrication approximations. In addition the forces depend on fluid velocities and particle velocities as independent variables. Practical implementation of this general framework in computer codes for many particles is still a great challenge, however. Commonly, the fluid velocity unknowns are eliminated by assuming that the fluid comoves with the particles at large scales \cite{bossis1984dynamic,brady1985rheology,Durlofsky1987,Ladd1989,Ladd1993,Cichocki1995,ball1997simulation}.
This assumption also entails that the pair drag forces must be frame invariant \cite{ball1997simulation}, which exclude many components of the full resistance matrices. In fact, all the long range interactions must be removed, and only the lubrication terms are left.

The assumption of co-movement is acceptable for shear flow at constant volume (type I), but it is otherwise a severe limitation of the numerical technique. One consequence of this assumption is that the solid fraction must be constant over space and time. It is not uncommon in practical applications to violate this condition. This is the case namely in sedimentation, or when the flow of a suspension has a free surface (e.g. debris flow or sheet flow), or when the shear occurs at imposed confining stress (type II experiments). In such case, the divergence of the large scale velocity fields of both phases balance each other, giving rise to long range hydromechanical coupling, also known as \emph{poromechanical} couplings in porous media theory and soil mechanics \cite{coussy2004poromechanics}. This strong coupling governs a range of phenomena such as liquefaction of loose materials or, conversely, delays in the solid-fluid transition in dense materials \cite{pailha2009two}. Note also that, even if both phases comove at large scales, the implicit fluid velocity field corresponding to the lubrication terms is not divergence free, and therefore not fully consistent.

Beside Stokesian Dynamics, numerical methods solving the Stokes or Navier-Stokes equations directly for the interstitial fluid have been introduced \cite{ladd2001lattice,yeo2010simulation}. In all cases, the so-called lubrication terms are singular and can not be solved explicitly by computational fluid dynamics (CFD). Capturing the divergence of these terms when particles approach contact would need to shrink the mesh to unrealistically small element sizes around the contact region. A practical approach is to let CFD compute the non-singular terms and to add the lubrication terms directly using analytical expressions \cite{Nguyen2002,Ding2003}.

In the present work, dense suspensions are simulated, using a particle-based method and accounting for three effects: contact interactions, drag forces resulting from the poromechanical coupling, and lubrication forces. The contact forces and the motion of the particles are computed using a Discrete Element Method (DEM). The poromechanical coupling is accounted for using the pore-scale coupling DEM-PFV (Pore Finite Volume) developed recently \cite{chareyre2012pore,Catalano2013}. The DEM-PVF code has been extended to periodic boundary problems for the purpose of the present study. And finally the lubrication forces are introduced using frame invariant expressions. 

One question that we will examine in this paper is whether accounting for the divergence free nature of fluid velocity field at the local scale can significantly affect the rheology of dense suspensions at constant volume. The poromechanical coupling will be exhibited as a transient effect during volume changes. Another question of interest concerns the definition of the lubrication forces and torques. As particles move one relative to one another, normal, tangential, rolling and twisting motions generate different effects. These effects are sometimes introduced selectively in numerical models, considering that some of them must have negligible effects, a priori. Typically, only normal and shear forces are computed \cite{rognon2011flowing,Nguyen2002,trulsson2012transition}, and sometimes even the shear force is neglected \cite{lemaitre2009dry,ancey1999theoretical,seto2013discontinuous}. Hereafter, we introduce all possible terms in order to evaluate, a posteriori, which ones can be neglected. Finally, the 
contributions of - respectively - the solid contacts, the lubrication forces, and the poromechanical coupling will be investigated.

\section{Numerical model}
\subsection{Discrete Element Model}
\label{DEM}
An explicit finite difference scheme is employed for updating the position of each particle in a time-marching algorithm. The particles move according to the Newton's second law. In the absence of gravitational effects, the motion is driven by elastic-frictional contact forces defined using a soft contact approach classical in the DEM \cite{Cundall1979}. The contact parameters are the normal and shear stiffnesses $k_n$ and $k_s$, and the angle of contact friction $\phi$. The contact forces are supplemented hereafter with forces coming from the interstitial fluid. A three-dimensional implementation of the DEM as found in the open source software YADE is used herein. For more details about the implementation, please refer to \cite{vsmilauer2010yade}.

\subsection{DEM-PFV coupled model}
\label{couplage}
\begin{figure}[htp]
\centering
\includegraphics[width=0.5\columnwidth]{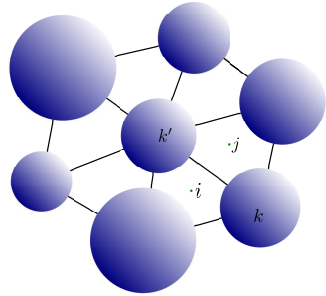}
\caption{Regular triangulation in 2D.}
\label{mesh}
\end{figure}

The DEM-PFV method is used to solve a pore-scale version of the mass balance equation which appears in the continuous theory of porous media and leads to the so-called poromechanical coupling.
Only the main steps of the method are outlined hereafter since the details can be found in previous papers. 
Here, we assume incompressible phases as in \cite{chareyre2012pore,Catalano2013} (for compressible phases see \cite{scholtes2015}). A tetrahedral decomposition of the pore space is introduced based on regular triangulation (figure \ref{mesh}), where that part of a tetrahedron occupied by the fluid is called a \textit{pore}. From now on $V_i$ denotes the volume of pore $i$. It is uniquely defined by the positions $\mathbf x_i$ and sizes of the solid particles, while the rate of change $\dot V_i$ also depends on their velocities $\dot {\mathbf x}_i$. 

An exchange of fluid between adjacent pores $i$ and $j$ is represented by the interface flux $q_{ij}$. The volume balance equation for one one pore leads to 
\begin{equation}
 \dot{V}_i\; = \; \sum_{j=1}^{j=4} \; q_{ij}.
\end{equation}
Assuming a Stokes regime entails a linear relation between $q_{ij}$ and the local pressure gradient $(p_i -  p_j)/l_{ij}$ between two pores, where $l_{ij}$ is a reference length (see \cite{chareyre2012pore}). It leads to the following relationship between pressure and rate of volume change:
\begin{equation}
\label{balance_eq}
 \dot{V}_i= \; \sum_{j=1}^{j=4} \; k_{ij} \; (p_j  -  p_i)/l_{ij} = \; \sum_{j=1}^{j=4} \; K_{ij} \; (p_j  -  p_i).
\end{equation}
In this equation $K_{ij}$ is the local hydraulic conductivity. It must reflects the small scale geometry of the packing. In details, the proposed expression of $K_{ij}$ depends on a local hydraulic radius $R^h_{ij}$ (area of the fluid-solid interface divided by the fluid volume - again see \cite{chareyre2012pore}) as
\begin{equation}
\label{permeability}
K_{ij}=\alpha \frac{S_{ij}^f {R_{ij}^h}^2}{2\eta\, l_{ij}}
\end{equation}
where $S^f_{ij}$ is the cross-sectional area of the pore-throat, $\eta$ is the viscosity of the fluid, and $\alpha$ can be interpreted as a calibration parameter. $\alpha=1$ is known to give good estimates of the actual permeability of glass beads \cite{Tong2012} but we used $\alpha<1$ in this study. This is further discussed in section \ref{sim}.

Substituting $\dot{V}_i$ by its expression in terms of particles velocity and writing equation \ref{balance_eq} for every element gives a system of linear equations. In a matrix form and including boundary conditions, the sytem reads 
\begin{equation}
\label{flow_solution2}
\mathbf{K} \mathbf{P} = \mathbf{E}\,\dot{\mathbf{x}}+ \mathbf{Q}_q+\mathbf{Q}_p,
\end{equation}
where $\mathbf{K}$ is the conductivity matrix containing the terms $K_{ij}$, $\mathbf{P}$ the column vector of pressure unknowns, and $\mathbf{E}$ is the matrix defining the rates of volume change of the elements such that $\dot{V_i}= (\mathbf{E}\,\dot{\mathbf{x}})_i$. $\mathbf{Q}_q$ and $\mathbf{Q}_p$ are flux vectors reflecting the boundary conditions, respectively source terms (imposed fluxes) in $\mathbf{Q}_q$ and imposed pressures in $\mathbf{Q}_p$.
Solved at each time step, equation \ref{flow_solution2} gives the discrete field of fluid pressure $\mathbf{P}$ as function of the particles velocity.

The drag forces can be deduced from the pressure field. They are the integrals of the pressure $p$ and the viscous stress $\tau$ on the surface of the particle (for a non-inertial fluid, the second integral can be evaluated based on $\mathbf{P}$ using momentum balance):
\begin{equation}
F^f_k  =  \int_{\partial \Gamma_k}  p \textbf{n}  ds +  \int_{\partial \Gamma_k}  \tau  \textbf{n}  ds 
\label{fporo}
\end{equation}
The forces $F^f_k$ are linearly dependant of $\mathbf{P}$, hence a matrix form giving the drag forces for all particles
\begin{equation}
\mathbf{F}^f = \mathbf I^f \mathbf P
\end{equation}

The $F^f_k$ are introduced in Newton's second law together with the forces coming from solid contacts ($\mathbf{F}^c$) and lubrication effects ($\mathbf{F}^L$ defined in the next section). I.e. 
\begin{equation}
\label{brute_newton_coupled}
\mathbf{M}\ddot{\mathbf{x}}=\mathbf{F}^c+\mathbf{F}^L+\mathbf I^f \mathbf P,
\end{equation}

The strong two-way coupling defined by equations \ref{flow_solution2} and \ref{brute_newton_coupled} is the poromechanical coupling. It is integrated with an explicit scheme whose accuracy has been verified in \cite{Catalano2013}.

\subsection{Lubrication}
\label{lubrication}
\begin{figure}
\includegraphics[width=\columnwidth]{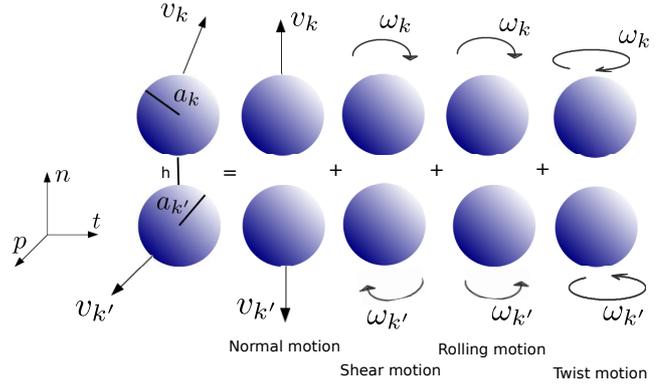}
\caption{Relative motion between particles.}
\label{motion}
\end{figure}

We assume that lubrication reflects the dominant dissipation process locally in the sheared pore fluid.
The lubrication effects are computed between particle pairs as soon as they share an edge in the triangulation (figure \ref{mesh}) (the triangulation is updated dynamically during the simulation). They are defined for all the elementary motions described in figure \ref{motion}. We consider particles $k$ and $k'$ with radii $a_k$ and $a_{k'}$, linear velocities $\bm v_k$ and $\bm v_{k'}$ and angular velocities $\bm\omega_k$ and $\bm\omega_{k'}$, respectively. Their average radius is defined as $a=(a_{k}+a_{k'})/2$ and $h$ denotes the inter-particle distance (surface to surface). An arbitrary relative motion between two particles can be decomposed in four elementary motions corresponding to normal displacement (subscript $n$), shear displacement ($s$), rolling ($r$) and twisting ($t$). This decomposition is illustrated in figure \ref{motion}. In addition, we introduce the angular velocity of the local frame attached to the interacting pair: $\bm\omega_n=(\bm v_{k'}-\bm v_k)\times \textbf{n}/(a_k+a_{k'}+h)$. Lubrication forces and torques induced by these elementary motions are defined as follow:

\begin{equation}
 \bm F_{n}^{L} \; = \; \frac{3}{2} \; \pi \; \eta \; \frac{a^2}{h} \; \textbf{v}_n
 \label{fn}
\end{equation}
\begin{equation}
\bm F_{s}^{L} \; = \; \frac{\pi \eta}{2} \; \left[-2a \; + \; (2a+h) \; ln\left(\frac{2a+h}{h}\right)\right] \; \textbf{v}_t
\label{ft}
\end{equation}
\begin{equation}
\bm C_{r}^{L} \; = \; \pi \; \eta \; a^3 \; \left(\frac{3}{2} \; ln \frac{a}{h}\; + \; \frac{63}{500} \; \frac{h}{a} \; ln \frac{a}{h}\right) \; \left[(\bm{\omega}_k - \bm{\omega}_{k'}) \times \textbf{n}\right]
\label{cr}
\end{equation}
\begin{equation}
\bm C_{t}^{L} \; = \; \pi \; \eta \; a^2 \; \frac{h}{a} \; ln \frac{a}{h}  \; \left[(\bm{\omega}_k - \bm{\omega}_{k'}) \cdot \textbf{n}\right] \; \textbf{n}
\label{ct}
\end{equation}
\\

where $\textbf{v}_n = ((\textbf{v}_{k'} - \textbf{v}_k) \cdot \textbf{n}) \; \textbf{n}$ is the normal relative velocity and $\textbf{v}_t = (a_k (\bm{\omega}_k-\bm{\omega}_n) + a_{k'}(\bm{\omega}_{k'}-\bm{\omega}_n))\times \textbf{n}$ is an objective expression of the tangential relative velocity. In this set of equations, the normal and shear forces, $\bm F_{n}$ and $\bm F_{s}$, are based on Frankel $\&$ Acrivos \cite{frankel1967viscosity,van1991modeling} whereas $C_{r}$ and $C_{t}$ are based on Jeffrey \& Onishi \cite{jeffrey1984forces,jeffrey1984calculation} - the reason of this choice will be discussed later. The total lubrication force $\bm F_k^{L}$ (resp. $\bm F_{k'}^{L}$) applied by particle $k'$ on particle $k$ (resp. by particle $k$ on particle $k'$) and the total torque $\bm C_k^{L}$ (resp.  $\bm C_{k'}^{L}$) applied by particle $k'$ on particle $k$ (resp. by particle $k$ on particle $k'$) relative to the particle center read:

\begin{equation}
\bm F_k^{L} \; = \; -\bm F_{k'}^{L} \; = \; \bm F_{n} \; +  \; \bm F_{s} , 
\end{equation}
\begin{equation}
\bm C_k^{L} \; = \;  (a_k + \frac{h}{2}) \;  \bm F_{s} + \; \bm C_{r} + \; \bm C_{t}, 
\end{equation}
\begin{equation}
\bm C_{k'}^{L} \; = \;  (a_{k'} + \frac{h}{2}) \; \bm F_{s} - \; \bm C_{r}  - \; \bm C_{t}.
\end{equation}

Note that $\bm F_{s}$ contributes to the total torques.

\begin{figure}[htp]
\hspace*{-3mm}
\includegraphics[width=\columnwidth]{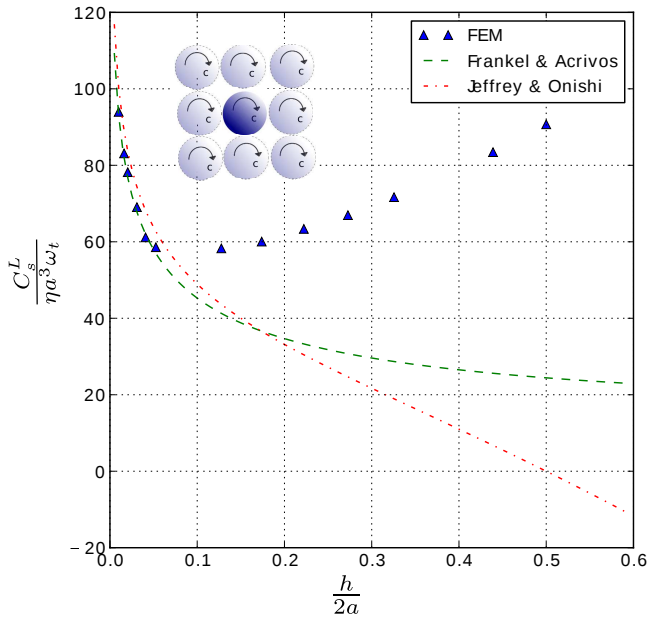}
\caption{Comparaison of viscous shear torques for the case of rotating sphere in a regular assembly of identical particles. h is the surface-to-surface distance and a is the particle's radius.}
\label{comsol}
\end{figure}

In order to check the validity of the different lubrication approximations for different interparticle distance $h$ 3D Finite Element Method (FEM) simulations of Stokes flow have been carried out. Periodic boundary conditions were used to represent an infinite array of identical spheres, fixed in space but all rotating at the same velocity (inset of figure \ref{comsol}). Figure \ref{comsol} presents the comparison of the FEM results with that of equation (9), where $F_{s}^{L}$ is determined alternatively using the expression from Jeffrey $\&$ Onishi \cite{jeffrey1984forces}:
$$ \bm F_{s} \: = \: \pi \: \eta \: a \: ln \frac{a}{h} \: \textbf{v}_t,$$\\
or from Frankel $\&$ Acrivos \cite{frankel1967viscosity}:
$$\bm F_{s}  \: = \: \frac{\pi \: \eta}{2} \: \left( -2 \: a \: + (2a+h) \: ln \frac{2a+h}{h}\right)  \: \textbf{v}_t.$$

Both expressions are asymptotically equivalent for $h\rightarrow0$. However, our results show that the second one is in much better agreement with the FEM result for small but finite distances ($h/2a<0.1$). Both expressions underestimate the FEM result for $h/2a>0.1$, which is not surprising keeping in mind that they have been obtained from asymptotic expansions. However, an additional defect of the former is that it leads to negative torques (i.e. torques with the same sign as the angular velocity) for large $h$. This feature is unphysical as it leads to a net creation of energy and can severely alter the stability of the numerical scheme. It was thus concluded that the expressions of Frankel $\&$ Acrivos were more suitable for implementation. 

\begin{figure}[htp]
\centering
\includegraphics[width=0.7\columnwidth]{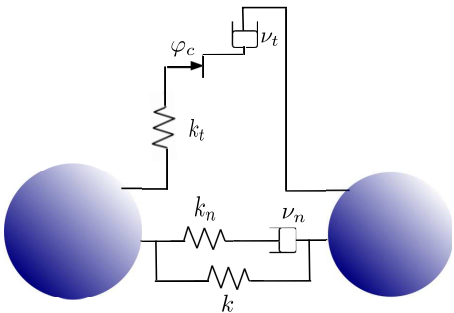}
\caption{Visco-elastic scheme of the interaction between two elastic-like particles.}
\label{maxwell}
\end{figure}

We account for the deformability of the particles near the contact region by combining the above normal lubrication model with a linear elastic model via a Maxwell-type visco-elastic scheme (\cite{Marzougui2013Part}, partly inspired by \cite{rognon2010}) (figure \ref{maxwell}). The parameters are $k_n$ the contact stiffness and $\nu_n(h)$ is the instantaneous viscosity of the interaction as defined in eq. (3), such that $F_{n}^{L}  = \nu_n(h) {v}_n$. The real velocity of approach between the two surfaces is $\dot h=v_n - \dot u_n^e$, where $u_n^e$ is the elastic deformation given by $u_n^e=F_n^L/k_n$. Hence, the evolution of the normal lubrication force obeys the differential equation 
\begin{equation}
F_n^L = \nu_n(h) \left( v_n - \frac{\dot{F_n^L}}{k_n} \right),
\end{equation}
that we have to integrate over time-steps using the form
\begin{equation}
\dot F_n^L = k_n \left( v_n - \frac{F_n^L}{\nu_n(h)} \right).
\end{equation}
The same stiffness $k_n$ is used for both the DEM contact model and the visco-elastic lubrication model.

\begin{figure}[htp]
\centering
\includegraphics[width=\columnwidth]{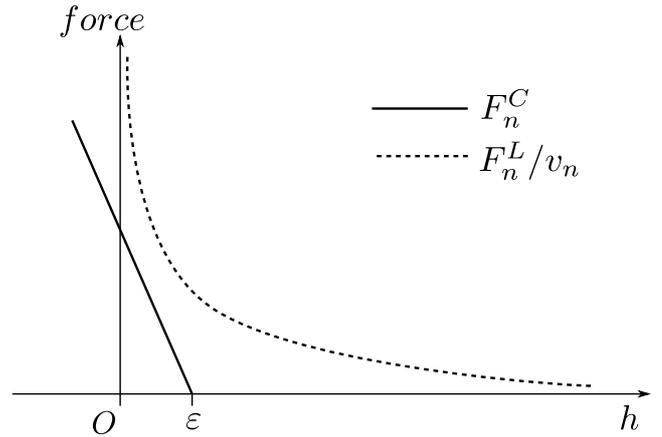}
\caption{Evolution of the contact force and of the normalized lubrication force in the normal direction as a function of the gap between two particles.
$\varepsilon$ defines the roughess of the particles surfaces.}
\label{roughness}
\end{figure}

Lastly, when the particles approach each other, the pressure in the gap tends to infinity and the normal lubrication force would theoretically prevent contact. Like many others (see e.g. \cite{rognon2011flowing}), we assume that contact will actually occur when $h$ is of the order of the particle roughness $\varepsilon$ (figure \ref{roughness}). Thus the solid contact model will generate a repulsive interaction even before $h=0$. As a result of this combination between a visco-elastic model and roughness at contact, $h$ will never reach 0 practically in numerical simulations.

\section{Numerical simulations}
\label{sim}
\subsection{Simulation setup}
\begin{figure}[htp]
\centering
\includegraphics[width=0.9\columnwidth]{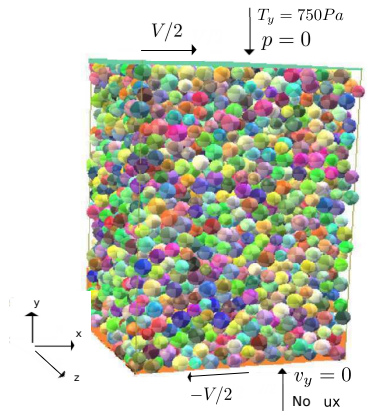}
\caption{Simulation cell.}
\label{sample}
\end{figure}

The suspension is represented by a bi-periodic packing made of $N=1000$ frictional spheres of average radius $a=0.025 \pm 0.01$ m. The physical properties are (unless stated otherwise for sensitivity analysis) roughness $\varepsilon=0.035\:a$, density $\rho = 2500$ kg/m$^3$, normal contact stiffness $k_n/a=5\times10^5$ Pa, shear stiffness $k_s=k_n/2$, and contact friction angle $\varphi = 30^{\circ}$. There is no gravity. The numerical sample is $H=18a$ high, $L=12a$ long and $l=12a$ wide (figure \ref{sample}). It is first confined between two parallel plates then sheared by moving the top and the bottom plates at constant velocity $\pm V/2 = 1.5$ m/s. The boundary conditions for the top plate are the velocities $v_x = V/2$, $v_z =0$, the total vertical stress ${T}_y = 750$ Pa and the fluid pressure $p=0$. At the bottom plate, $v_x = -V/2$, $v_z = 0$ and the fluid velocity along the y axis $v_y^f = 0$ (impermeable boundary). Periodic boundary conditions are defined along the horizontal axis for both the particles and the fluid. For the latest we impose a null pressure gradient at the macro-scale, i.e. $\bigtriangledown p_x = \bigtriangledown p_z =0$  (see \cite{Marzougui2013Part}). In order to avoid preferential slip zones near the plates, the first layer of spheres in contact with a plate is fixed to the plates by highly cohesive contacts.
We introduce the boundary stress vector $\bm{T} = \bm{F} / S$ where $\bm{F}$ is the total force on the top plate and $S$ is the horizontal cross sectional area. ${T}_y$ is constant during the deformation, while ${T}_x$ is a result of the imposed shear.

\subsection{Transient vs. steady state}
Figure \ref{brut} shows the evolution of the shear stress $T_x$, the pressure $p$ and the solid fraction $\phi$ function of the deformation $\gamma(T) = \int_0^T \dot{\gamma}(t)dt$ where $\dot{\gamma}(t) = V / H(t)$ is the shear rate. The numerical results are presented for two cases: a first case where the poromechanical coupling is considered and a second one without it (i.e. ignoring the last term in equation \ref{brute_newton_coupled}). When the shear velocity is applied, a transient regime is observed, characterized by an increase followed by a decrease of the shear stress, a decrease of the solid fraction and a negative pressure of the fluid in the coupled case. This later effect entails a higher effective stress in the coupled problem, explaining why the shear stress reaches higher values. The system evolves toward a steady state for large deformations, in which the shear stress and the solid fraction are approximately constant and the pressure is nearly zero. The poromechanical coupling has no visible effect at steady state: the shear stress and the solid fraction reach similar values for both cases.

It is to be noted that the poromechanical coupling entails long range effects in the system and, ultimately, a dependency on the problem size ($H$ in our case).
It is known since Terzaghi that the characteristic time of such process scales with $\eta H^2/\kappa$ where $\kappa$ is the intrinsic permeability. Since $\kappa$ scales with $a^2$, the relaxation time of the transient regime is proportional to $\eta(H/a)^2$. A consequence is that the peak pore pressure in figure \ref{brut} scales with $(H/a)^2$, pressure gradients scale with 
$H/a^2$, and finally drag forces scale with $H$. Important consequences of this feature are that the drag forces and the lubrication forces are not comensurable, and that poromechanical effects cannot be reflected as rheological properties of the bulk material - it is always necessary to solve a coupled problem.

Since our main focus in this study was the bulk viscosity of suspensions (a rheological property), we did not seek a fully realistic combination of drag forces and lubrication forces. Practically, it let us reduce the computation times tremendeously by setting $\alpha=100$ in equation \ref{permeability}, while it should be close to 1 for more realistic simulations. The duration of the transient regime would have been multiplied by 100 approximately, and the timesteps of the time marching algorithm would have been reduced by 100, leading to an increase of the total simulation time by a factor $10^4$. Though qualitatively correct, the trends seen in fig. \ref{brut} are thus quantitatively wrong (and remember that in any case they are only relevant for a specific value $H$).


\subsection{Stress decomposition}
\begin{figure}[htp]
\includegraphics[width=\columnwidth]{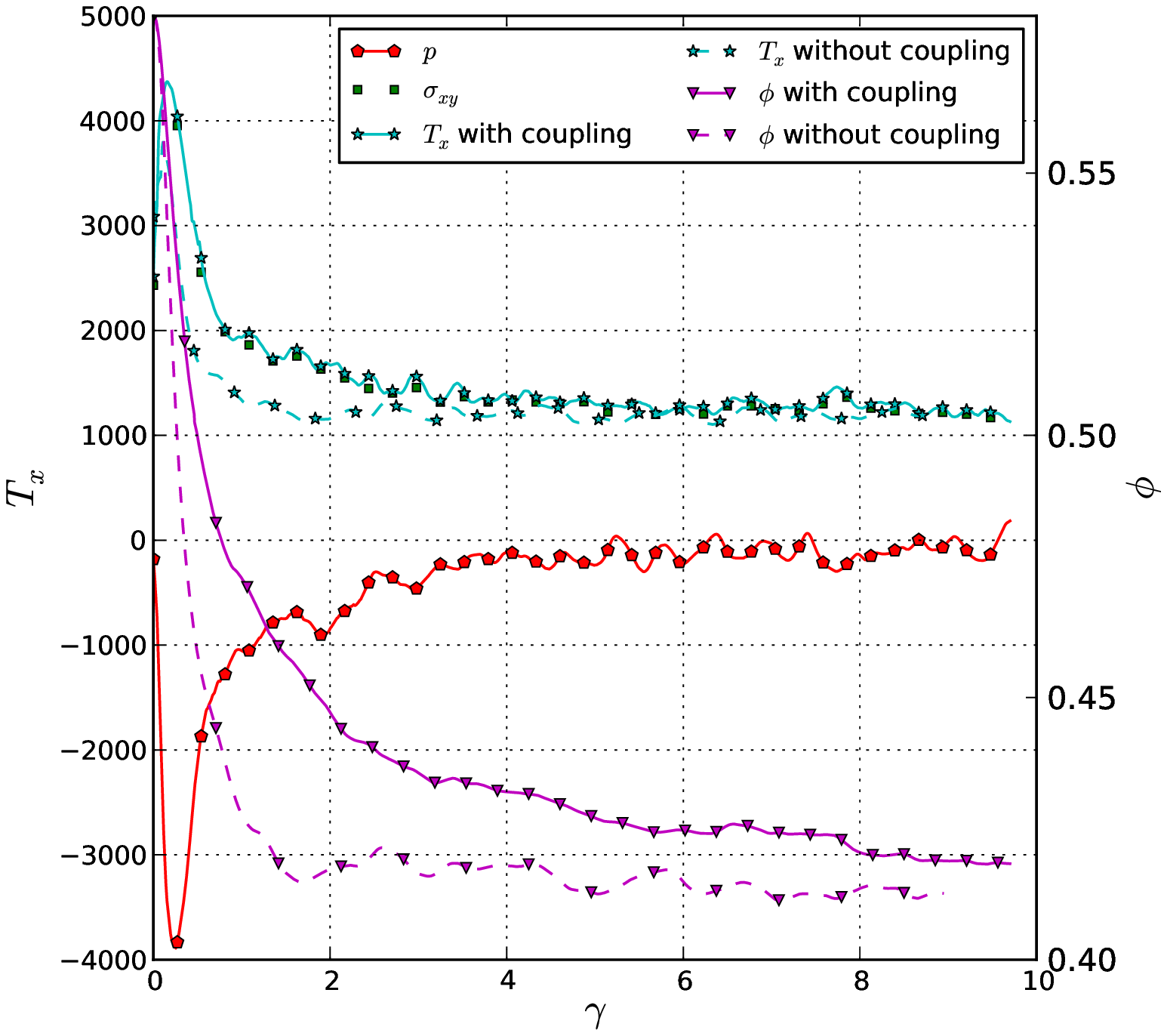}\newline
\newline
\includegraphics[width=0.88\columnwidth]{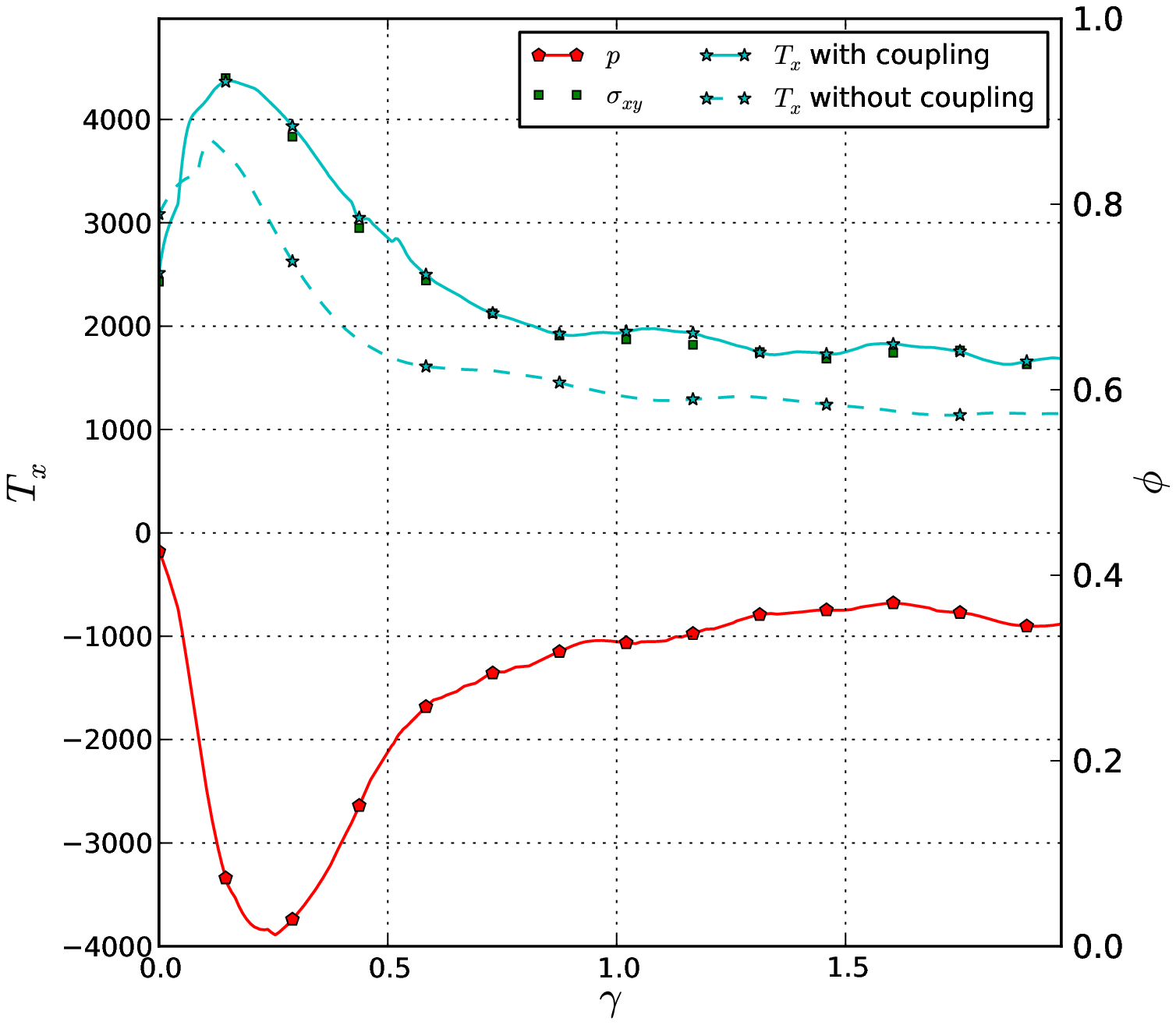}\newline
\caption{(top): The evolution of the shear stress and the solid fraction as a function of the deformation for two cases: with poromechanical coupling and without poromechanical coupling. (bottom): Zoom on the transient regime.}
\label{brut}
\end{figure}

Besides $\bm{T}$, a tensor representing the average stress in the suspension can be calculated as
\begin{equation}
\bm{\sigma} \: = \: \bm{\sigma}^{C} \: + \: \bm{\sigma}^{L}  \: + \: p \textbf{I} + \: \bm{\sigma}^{I}
\label{decomp}
\end{equation}
$\bm{\sigma}^{C} = \frac{1}{V} \sum_{ij} \bm{F}^{C}_{ij} \otimes \bm l_{ij}$ is the contact stress applied on particles in contact where $\bm l_{ij}$ denotes the branch vector between the centers of the particles $i$ and $j$. Similarly, $\bm{\sigma}^{L} = \frac{1}{V} \sum_{ij} \bm{F}^{L}_{ij} \otimes \bm{l_{ij}}$ is the contribution from lubrication forces \cite{ancey1999theoretical}, which will be further decomposed hereafter by considering separately the normal and shear components of the lubrication force. $p$ is the pressure associated to the poromechanical coupling \cite{Catalano2013}.  $\bm{\sigma}^{I}=\sum_k m_k \textbf{v}_k \otimes \textbf{v}_k$ reflect the inertial effects as defined in \cite{savage1981stress} where $m_k$ is the mass of particle $k$ and $\bm v_k$ is its velocity. It can be verified in figure \ref{brut} - where both $T_x$ and $\sigma_{xy}$ are plotted - that the two expressions compare consistently. Hereafter, eq. (\ref{decomp}) will be used to assess the microscale origins of the shear stress and their rate dependency. 

\section{Results and Discussion}
\label{res}
\subsection{Typical results}
\label{resTypical}
\begin{figure}
\includegraphics[width=\columnwidth]{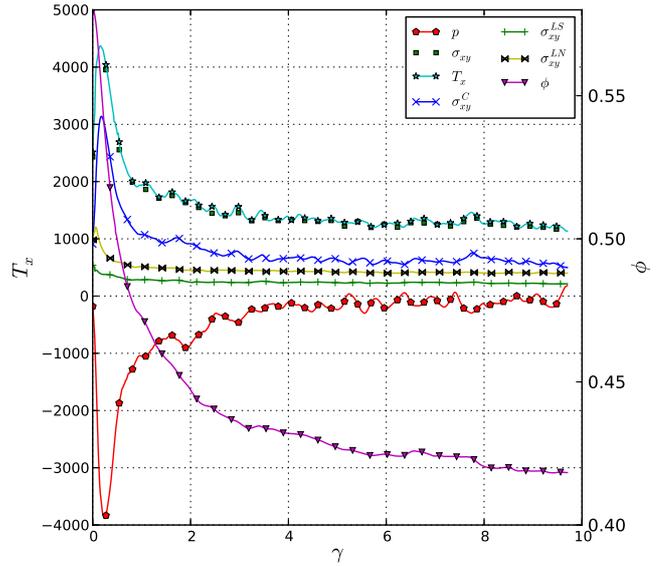}
\caption{Decomposition of the total shear stress in contact stress, normal lubrication stress and shear lubrication stress.}  
\label{resultStress}
\end{figure}

Figure \ref{resultStress} includes the evolution of different terms of equation (\ref{decomp}) (component $xy$) as the suspension is sheared at $I_v = 0.21$. The inertial stress $\sigma^I_{xy}$ (not represented here) is negligible compared with the total stress ($\sigma^{I}_{xy} < 2.5 \% \: T_x$), which indicates that the suspension is dominated by contacts and viscous interactions in this case (further discussion in the next paragraph). Second, the contact stress contributes to approximately half of the total stress ($\sigma^{C}_{xy} \approx 50\% \: T_x$) whereas the other half is due to the normal and shear lubrication forces ($\sigma^{LN}_{xy} \approx 30\% \: T_x$  and $\sigma^{LS}_{xy} \approx 20\% \: T_x$). The different contributions will be further investigated for different values of the viscous number $I_v$ in the following.

\subsection{Viscous number}
$I_v$ is a dimensionless form of the shear rate \cite{boyer2011unifying}, reflecting the magnitude of viscous effects, and is defined as: 
\begin{equation}
I_v = \frac{\eta \; \big\vert \dot{\gamma} \big\vert  }{T_{y}},
\end{equation}

\begin{figure}
\includegraphics[width=\columnwidth]{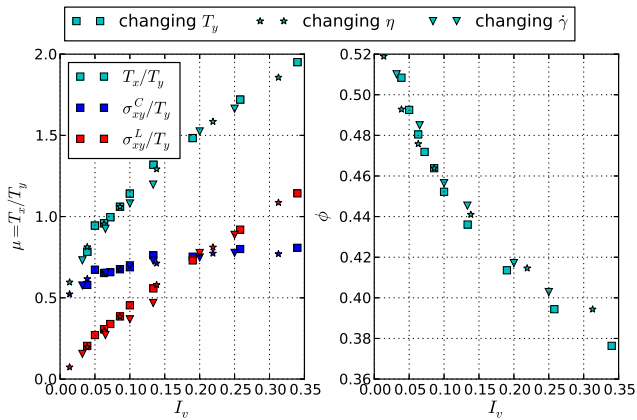}
\caption{Normalized shear stress and solid fraction at steady state versus $I_v$. In each series the change of $I_v$ is obtained by changing a different parameter: confining pressure ($\blacksquare$), viscosity ($\bigstar$), or shear rate ($\blacktriangledown$).}  
\label{resultStress2}
\end{figure}

The key idea here is that, in the viscous regime and at steady state the stress ratio $\mu = T_x / T_{y}$ and the solid fraction $\phi$ are entirely controlled by this unique parameter. In other words, all possible combinations of confining pressure, fluid viscosity, and shear rate corresponding to a given value of $I_v$ should give the same result. In order to confirm this property, three series of simulations were conducted in which the control parameters were changed independently in each series to produce different values of $I_v$. The results are plotted in fig. \ref{resultStress2} versus the corresponding values of $I_v$. $\mu$ and $\Phi$ are nearly the same whatever the method to change $I_v$. This result offers a numerical confirmation of the conclusion of Boyer et al. \cite{boyer2011unifying}.
This property holds only in the viscous regime, i.e. as long as the inertial effects can be neglected. As suggested in \cite{trulsson2012transition}, this condition may be characterized by the ratio $I/I_v$, where $I=\sqrt{\rho \dot \gamma^2/T_y}$ is the so called inertial number. Here the ratio is at most $I/I_v=5$, and it is much smaller in most cases, while \cite{trulsson2012transition} suggest $I/I_v\simeq10$ for the transition from the viscous to the inertial regime. For the rest of this study it was decided to keep $I$ constant and equal to 0.14, which leaves $\eta$ as the only free parameter. An exception is the dry case ($I_v=0$) where $I=0.005$.

\subsection{Dropping terms of the hydrodynamic model}
The consequences of neglecting some of the terms defined by eqs. (\ref{fporo})-(\ref{ct}) will be examined, in terms of stress ratio $\mu$ and of solid fraction $\phi$. This question is of interest for the developpers of numerical models, since many models found in the literature are not including all terms. As we have seen before, neglecting the poromechanical coupling (eq. \ref{fporo}) has a detrimental effect on the transient state but has no strong effect at steady state. We now focus on the different lubrication terms and how they affect the result at steady state.

\begin{figure}
\includegraphics[width=\columnwidth]{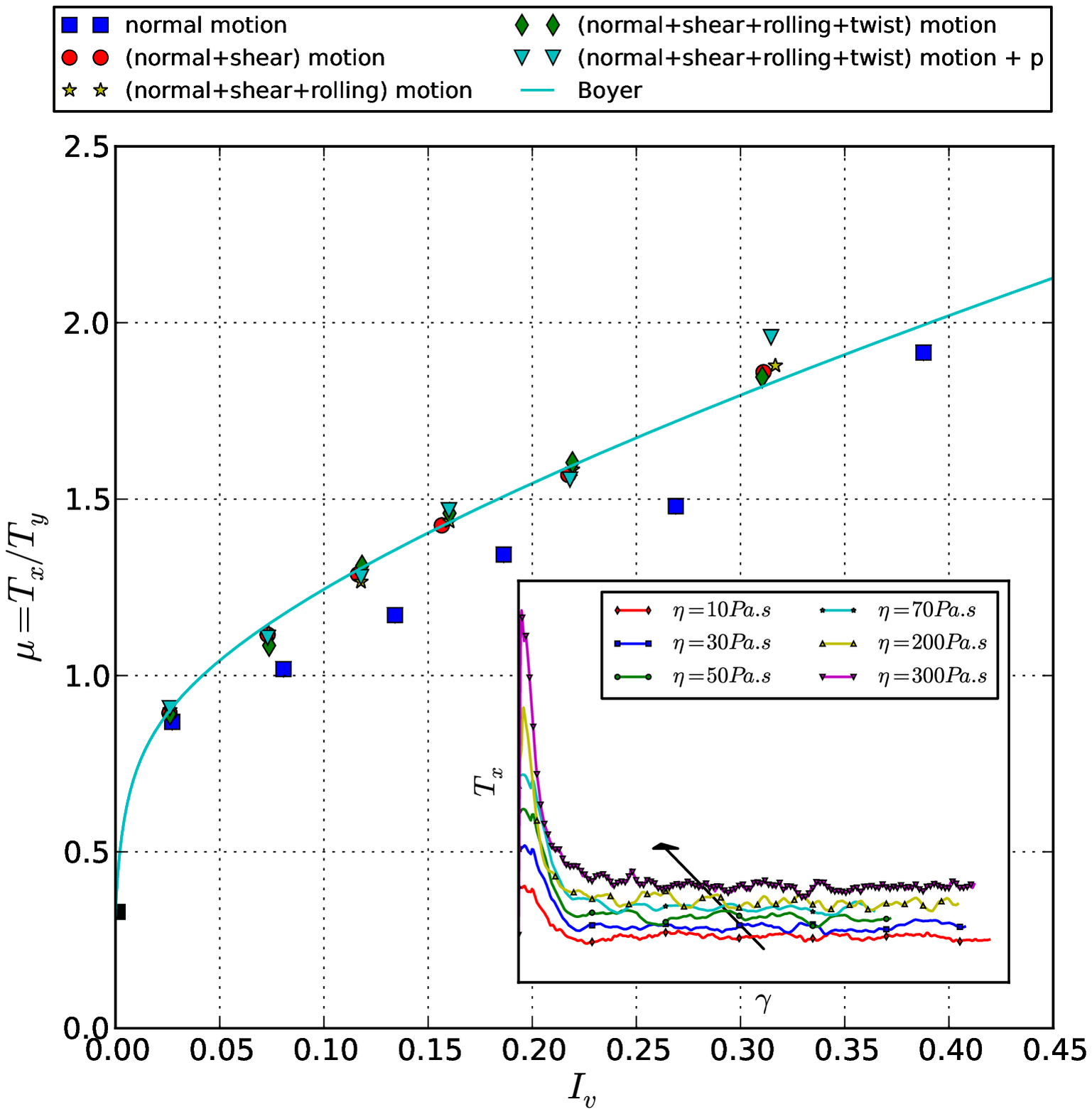}\newline
\newline
\newline
\includegraphics[width=\columnwidth]{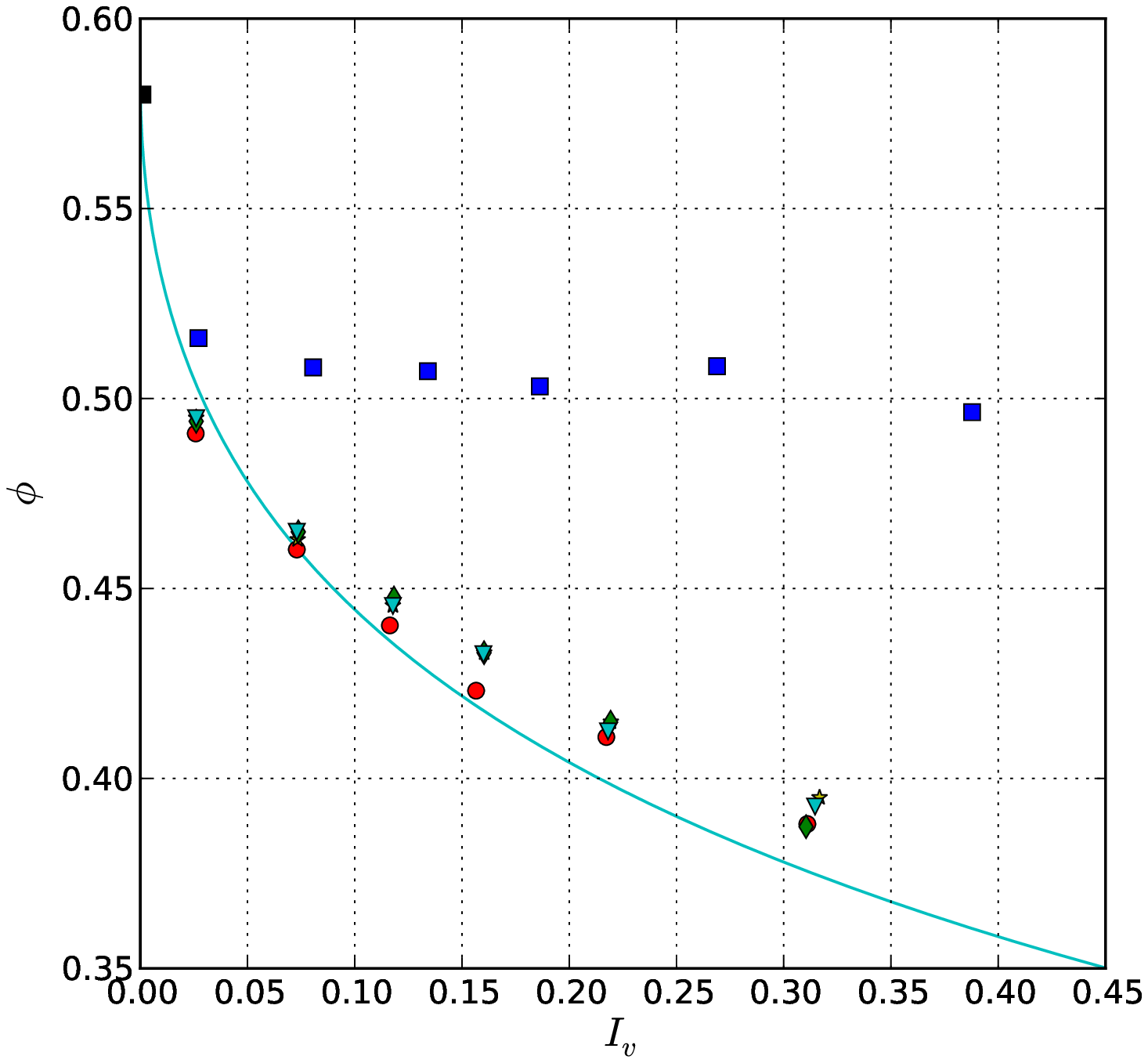}\newline
\caption{Stress ratio $\mu$ (top) and the solid fraction $\phi$ (bottom) at steady state versus $I_v$. Each symbol represents a different combination of lubrication terms.
The solid line is the phenomenological law of Boyer et al. \cite{boyer2011unifying}. Inset: the total shear stress for different values of fluid viscosity.}
\label{mu}
\end{figure}

Figure \ref{mu} presents the comparison of the numerical results with the phenomenological laws proposed by Boyer \textit{et al} \cite{boyer2011unifying} on the basis of experiments. The values reported in this figure have been obtained at steady state by increasing the fluid viscosity while keeping the shear rate $\dot{\gamma}$ and the normal stress $T_{y}$ constant. Typical results for $T_{x}$ vs. $\gamma$ are shown in the inset of the figure for different values of $\eta$. Series of simulations have been carried out including different combinations of the hydrodynamic effects. Starting with the simplest case where only solid contacts and normal lubrication forces are present, the other hydrodynamic terms are added one by one: shear lubrication force, rolling torque, twist 
torque and drag forces due to the poromechanical coupling.

The result for $I_v=0$ corresponds to vanishing hydrodynamic forces. It is obviously independent of the hydrodynamic assumptions, and practically it is the result of a simulation for a dry material. The friction coefficient and the solid fraction obtained in this case closely match the values measured experimentally by Boyer et al.
The results obtained with normal lubrication forces only (blue squares) match at least qualitatively the empirical evolution of shear stress with $I_v$. It is to be noted however that the solid fraction obtained with this model is almost constant for $I_v > 0.02$, while the experiments suggest a monotonic decrease. As soon as the shear lubrication forces are included, the results (red circles) get closer to the phenomenological law for both $\mu(I_v)$ and $\phi(I_v)$. It can be concluded that shear lubrication forces play a key role in the rate dependent dilatancy, and they contribute significantly to the shear stress. The normal lubrication alone lead to a satisfactory stress ratio but significantly overestimate the solid fraction. Considering this result, one may expect that simulations at imposed volume (called type I in the introduction) and including only normal lubrication forces would underestimate the shear stress even more than in our case.

Further sophisticating the hydrodynamic model does not yield other significant changes. The rolling torques (yellow stars), the twist torques (green diamond) and the poromechanical coupling (cyan triangles) have only marginal effects. We recall that this conclusion holds at the steady state only. As discussed previously, the poromechanical coupling can significantly interfere in the transient regimes.

\subsection{Contact stress versus hydrodynamic stress}
\begin{figure}
\includegraphics[width=\columnwidth]{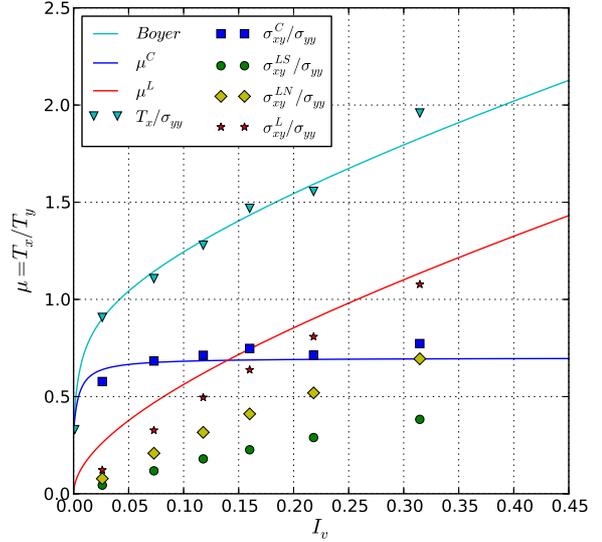}
\caption{The stress ratio $\mu$ and the decomposition in contact stress, the normal lubrication stress and the  shear lubrication stress. The solid line is the phenomenological law of \cite{boyer2011unifying}.}
\label{inset}
\end{figure}

From now on, the results that will be analyzed are obtained with the full model including all possible hydrodynamic interactions. Figure \ref{inset} shows the contribution to the total stress of the different terms in eq. (\ref{decomp}). The contact forces play a significant role for all the values of $I_v$ investigated. The contact stress slightly increases for $0<I_v<0.1$ and saturates to an almost constant value for larger values of $I_v$. Lubrication stresses, both normal and shear components, increase almost linearly, and the shear stress due to the shear components is approximately twice smaller than the stress coming from the normal components. For values of $I_v\ge 0.15$ the sum of the two lubrication stresses exceeds the contact stress. This result highlights the fact that depending on the value of $I_v$ two regimes are observed. At low $I_v$ the contact interactions are dominant whereas for $I_v\ge 0.15$ the lubrication interactions dominate. 
This result confirms a constitutive property inferred by Boyer \textit{et al} \cite{boyer2011unifying}. Herein, the shear stress in dense suspensions is split into two contributions, one coming from the contacts and represented by the same phenomenological law as in dry granular flow, the other one coming from hydrodynamic interactions similar to a Krieger-Dougherty viscosity. However Boyer \textit{et al} inferred this stress partition from macroscopic measurements. The present results allow to further assess the respective contributions of contacts and hydrodynamic interactions. Also, the choice of a frictional rheology for the contact stress which saturates for high values of $I_v$ is predicted by our discrete numerical simulations. We believe that this property is not trivial. Based on solid fraction at the largest $I_v$ indeed ($\phi\simeq 0.38$, far below any values that can be reached in dry quasi-static granular systems), one could expect that no solid contacts persist. Nevertheless, 
the numerical simulations confirm the proposal of Boyer \textit{et al.} on this aspect.

\subsection{Roughness}
\begin{figure}
\includegraphics[width=\columnwidth]{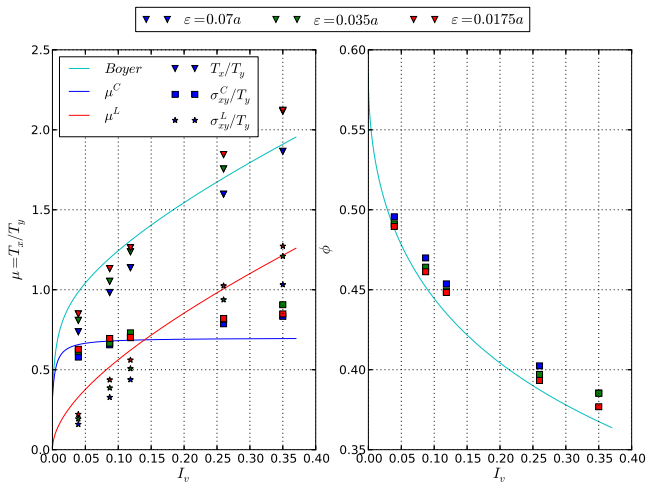}
\caption{Stress ratio $\mu$ and solid fraction $\phi$ at steady state for different values of the roughness parameter $\epsilon$.}
\label{roughnessEffect}
\end{figure}

The particles roughness $\epsilon$ appears as a key parameter in the lubrication model since the surface-to-surface distance $h$ may vanish as $\epsilon \rightarrow 0$, a situation in wich the lubrication terms diverge. This situation is peculiar from a physical point of view but it also causes major troubles from a numerical point of view. Arguably perfectly smooth surfaces are rare and this parameter can be justified on a physical ground. It is not always clear how this should be accounted for in models however, and in our case we don't have a precise knowledge of what value of $\epsilon$ would be relevant for the spheres used by Boyer et al. In order to evaluate the role of this parameter simulations were reproduced for three different values of $\epsilon$. They are reported in figure \ref{roughnessEffect}, which includes the total stress and the contributions from contacts and lubrication forces.

The lesser the roughness, the larger the contribution of lubrication, as one could expect. A less expected result is that the contribution of contacts is not strongly modified. No clear trend can be distinguished, as the points corresponding to the smallest $\epsilon$ may be below or above the others depending on the value $I_v$. This suggests that the different values are simply due to imprecisions in the evaluation of the steady state, and that the contribution of contacts could be independent of roughness. Overall, the difference on the total stress is of the order of 10\% or less, showing a relatively moderate effect. An effect on dilatancy is visible, the smoother particles dilate more, but again moderate. All three values seem to reflect the main features of the behaviour in spite of quantitative differences. Roughness does not appear as a key parameter here. The limit $\epsilon \rightarrow 0$ remains as an open question, which can't be studied easily with our method due to numerical difficulties.

\subsection{Microstructure}
\begin{figure*}
\includegraphics[width=\textwidth]{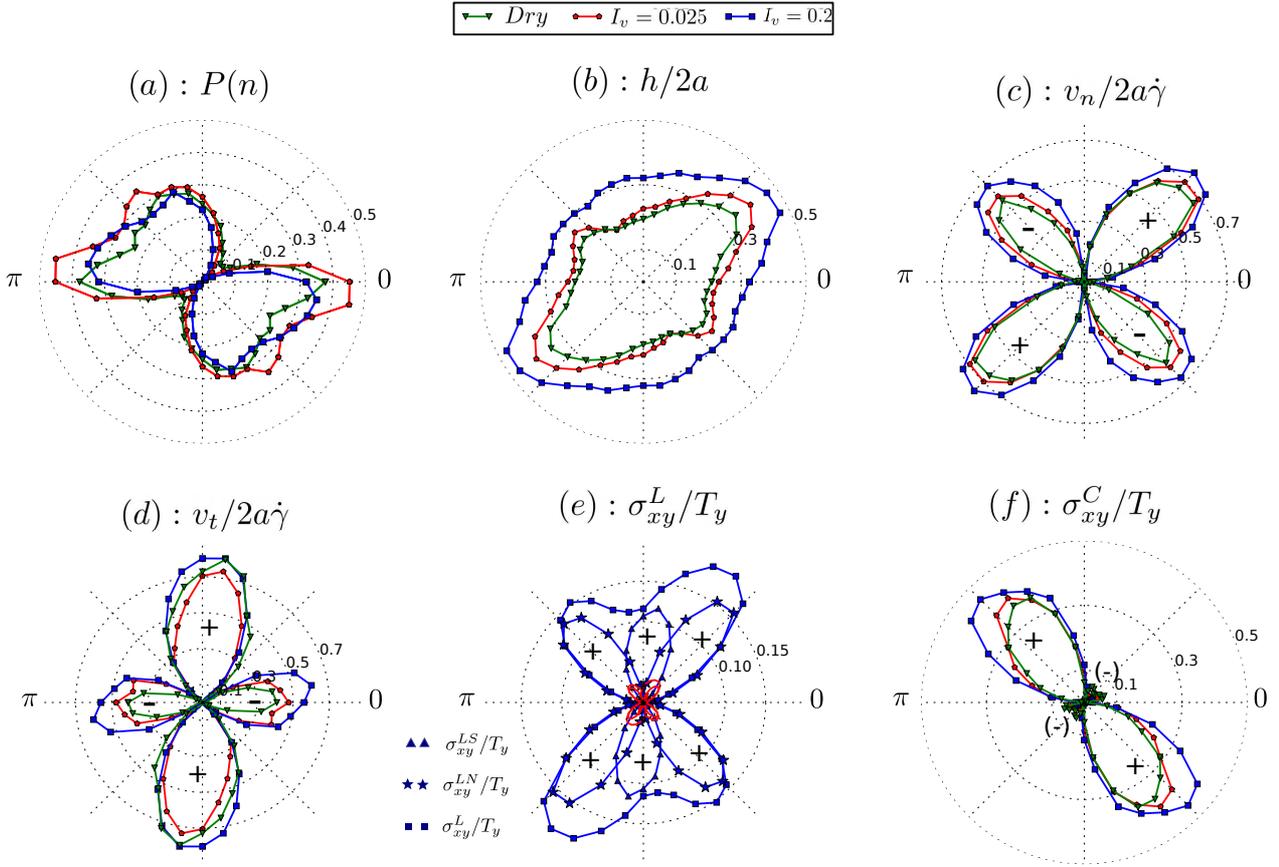}
\caption{Distributions of  normalized quantities in the (x,y) plane [top/left] for the dry case, $I_v = 0.025$ and $I_v = 0.2$: PDF of contact orientation, normal velocity, shear velocity, lubrication stress which is the sum of the normal and shear lubrication stress, and contact stress.}
\label{distribution}
\end{figure*}

In order to link averaged quantities to micro-scale variables, we examine how various quantities depends on the orientation of the particle pairs (figure \ref{distribution}). Based on the orientation of the unit normal, every interaction corresponds to a position on the unit sphere. For an arbitrary point $M$ on the unit sphere it is possible to compute averages of interaction variables. For instance, the average distance between the spheres is defined as: $\overline{h}(M)= \sum_{\bm{n}_k \in dS} h(k)/N_M$ where $dS$ is a small angular sector centered on $M$ and $N_M$ is the number of interactions associated to $dS$. In figure \ref{distribution} this value is normalized by the particle diameter ($2a$). We define similarly the average normal velocity and average shear velocity normalized by $2a\dot \gamma$.

The other plots of figure \ref{distribution} are density functions. The density of contacts is obtained by counting the number $N^c$ of interactions with $h<\epsilon$ in $dS$, and dividing by the total number of spheres $N^s$, so that $P(M)=N^c(M)/(dS\,N^s)$. The densities of the lubrication stress term $\bm{\sigma}^{L}(M)$ and of the contact stress term $\bm{\sigma}^{C}(M)$ are obtained by restricting the sums defined for eq. (\ref{decomp}) to the subset of interactions associated to $M$. For instance $\bm{\sigma}^{L}(M) = \frac{1}{V\,dS} \sum_{n_{ij} \in dS} \bm{F}^{L}_{ij} \otimes \bm{l_{ij}}$. On figure \ref{distribution} are the component $xy$ of both stress tensors, normalized by the confining pressure. The lubrication stress is further decomposed into one part due to normal forces and another part due to shear forces.

Despite the fact that all the functions introduced above define surfaces in the 3D space, in figure \ref{distribution} only the values for $M$ in the $(Oxy)$ plane  are plotted. The results are given for three different values of $I_v$: $I_v = 0$ (\textit{i.e.} a dry medium - green line in the figure), $I_v = 0.025$ (red line) and $I_v = 0.2$ (blue line). All distributions are $\pi$-periodic, the comments hereafter refer to the interval $[0,\pi]$.

%

Figure \ref{distribution}.a shows that there is a minimum in the density of contacts near $\theta = \pi/4$ for all cases. Conversely, the higher density is observed for orientations between $3\pi/4$ and $\pi$. A noticeable effect of increasing $I_v$ is that the number of contacts near $\theta = \pi/4$ vanish. This peculiar effect makes a clear difference between the viscous regime and the dry regime as $P(\theta = \pi/4)$ is always strictly positive in the latest. 

The average distance (figure \ref{distribution}.b) is increasing with $I_v$ on overall. Since increasing $I_v$ corresponds to a decrease of solid fraction, this trend is not surprising. The average distance is anisotropic and takes larger values near $\theta = \pi/4$, consistently with the lower density of contacts in this region.

The normal component of normalized relative velocity (figure \ref{distribution}.c) is positive on $[0,\pi/2]$ and negative on $[\pi/2,\pi]$. The extrema are of the same order in both cases, although it can be noted that the velocity of approaching particles ($[\pi/2,\pi]$) is slightly lower. The shear component (figure \ref{distribution}.d) is positive on - approximately - $[\pi/4,3\pi/4]$ and negative on $[0,\pi/4]\cup[3\pi/4,\pi]$. In this case the extrema are clearly different, with the largest relative velocities near $\theta = \pi/2$. We note that the graph of the shear velocity is not exactly symmetric as the maximum values are in fact a bit before $\theta = \pi/2$. It can be explained by smaller $h$ and more solid contacts preventing sliding when $\theta > \pi/2$.

The magnitude of the normalized relative velocity is increasing slightly with $I_v$. This is observed for both the normal and the shear components. This trend can be explained by considering the growing average distance between particles. If all particles were simply following the mean field velocity, then the relative velocity would obviously grow with $h$. Since the local fluctuations of velocity with respect to the mean field are not modifying the relative velocities in average, this correlation holds.

The lubrication forces for a given relative velocity are decreasing functions of $h$. Thus, it could be anticipated from figure \ref{distribution}.b that the contribution of viscous interactions to 
the bulk stress is dominated by interactions oriented along $\theta = 3\pi/4$, which have in average smaller values of $h$ but nearly similar values of normal velocity. The results of figure \ref{distribution}.e show a quite different picture. The density of stress due to the normal lubrication forces is actually slightly larger on $[0,\pi/2]$ and it matches the shape of the normal velocity closely (with the difference that the sign is always positive due to the branch $\bm l$ in the diadic product $F^L_n\otimes \bm l$). This feature may be explained by the strong correlations between $h$ and $v_n$. 

The contribution of the shear lubrication forces to the bulk stress is dominated by the interactions near $\theta = \pi/2$, consistently with the evolution of average shear velocity (the fact that the contribution vanishes when $\theta = 0$ is an effect of the product $F^L_s\otimes \bm l$ with $\bm l$ nearly horizontal). Like $v_s$, the density of stress reaches a pick slightly before $\theta = \pi/2$.

Considering the cumulated contributions $\bm \sigma^L =  \bm \sigma^L_n + \bm \sigma^L_s$, the non-symmetry of the density of stress becomes even more visible. The larger contribution is due to interactions in the range $[\pi/4,\pi/2]$. This is also the range in which there are nearly no solid contacts between the particles. Thus the non-symmetry may be due to a complex interplay between the solid contacts and the viscous interactions.

The density of contact stress (figure \ref{distribution}.f) is strongly anisotropic, as expected from the lack of contacts on $[0,\pi/2]$. When $I_v=0$ some contacts in $[0,\pi/2]$ contribute negatively, but the contribution is small. For $I_v>0$ this contribution becomes negligible. On overall, the contacts contribute to the bulk stress mainly via repulsive forces in the direction near $\theta = 3\pi/4$. The contribution of contacts to the bulk stress increases with $I_v$. This feature was already observed in figure \ref{mu} on the macroscopic variables.

 
\section{Conclusion}
We described a complete modeling framework for the numerical simulation of dense suspensions. The model includes solid contacts between particles using a soft contact approach, short range hydrodynamic interactions defined by frame-invariant expressions of forces and torques in the lubrication approximation, and the poromechanical coupling solved using the DEM-PFV technique.

Numerical experiments of simple shear at imposed confining normal stress have been conducted in an attempt to reproduce recent rheometer experiments on beads. The simulations are in excellent agreement with the empirical data in terms of bulk shear stress and solid fraction at the steady state, for a range of dimensionless shear rate corresponding to $0\leq I_v<0.45$.

The poromechanical coupling was shown to have a significant effect in the transient regime when the deformation starts. However, no significant effects of this coupling have been exhibited at steady state. 

The results obtained by neglecting some of the lubrication terms leads to the following conclusions. First, the normal lubrication term has the larger contribution to the bulk stress. However, considering this term alone leads to underestimate the shear stress, and with such simplification the model is unable to reflect the change of solid fraction for increasing $I_v$. Combining both normal and shear lubrication terms gives much better results in terms of stress and solid fraction. Further sophistication of the model by including the terms associated with rolling and twisting gives only marginal improvements.

The analysis of the various contributions to the bulk stress: contact forces, hydrodynamic forces and fluid pressure, has lead to the following conclusions. The contribution of contacts to the bulk shear stress in the permanent regime increases with $I_v$. This result may be seen as counter-intuitive. First, as higher $I_v$ leads to lower solid fraction, one would expect contacts contribution to decrease progressively with $I_v$ and ultimately vanish. Second, the assumption that the lubrication effects would prevent solid contacts in suspensions has been the cornerstone of many theoretical and numerical models in the past. Instead, the simulations reported in the present paper suggest that both the contact stress and the lubrication stress increase monotonically in the range of $I_v$ investigated.

As roughness of particules is decreased, the contribution of lubrication forces is increased to some extent. However, it does not lead to a decrease of contact forces, which remain nearly unchanged. Again, it is against the idea that lubrication is preventing solid contacts. The reasoning leading to this idea may be simply flawn due to conceptual mistakes. Namely, lubrication forces are often perceived as repulsive force wereas they are neutral overall, inhibiting the opening and the closure of contacts almost equally as revealed by microstructural variables.

The distribution of micro-structural variables revealed a complex interplay between the contact fabric and the hydrodynamic interactions. The anisotropy of contact orientation appears to be more pronounced in suspensions as compared to dry granular materials, due to the effect of the hydrodynamic interactions. This can explain at least partly why the contribution of contacts to the bulk shear stress is increasing.

This contact stress may reach a maximum and ultimately vanish for larger $I_v$. When does it happen remains an open question. Capturing this transition in numerical simulations is a great challenge for numerical models since the lubrication terms do not reflect appropriately all hydrodynamic interactions in more dilute regimes.

\section{Acknowledgments}
This work is supported by the PhD grant awarded to D. Marzougui by the University of Grenoble-Alpes.

\section{Disclosures}
This work has been supported by the PhD grant awarded to D. Marzougui by the University of Grenoble-Alpes. The content of the manuscript has not been published in previous publications. The study participants are the first author and co-authors and they consented to publish on June 12th, 2014. No institutional review board was required for this publication. The work as a whole has been approved by the IMEP-2 comitee of doctoral studies at Univ Grenoble Alpes on july 3rd 2011.


\end{document}